\documentclass{article}[11pt]  
\usepackage{amsmath}
\usepackage{amssymb}
\usepackage{amsthm}
\usepackage[letterpaper, margin=.75in]{geometry}
\usepackage{cite}
\usepackage{subcaption}
\usepackage{float}
\usepackage{enumitem}
\usepackage{graphicx}
\usepackage{multirow}
\usepackage{hyperref}
\usepackage{multirow}

\theoremstyle{definition}
\newtheorem{theorem}{Theorem}[section]
\newtheorem{lemma}[theorem]{Lemma}
\newtheorem{corollary}[theorem]{Corollary}

\newcommand{\para}[1]{\vspace*{.1cm}\noindent\textbf{#1.}}

\begin{document}

\title{A Simple Proof that Ricochet Robots is PSPACE-complete\thanks{This research was supported in part by National Science Foundation Grants CCF-1817602 and CCF-2329918.}}

\author{Jose Balanza-Martinez \and Angel A. Cantu \and Robert Schweller \and Tim Wylie}

\date{}
\clearpage\maketitle
\thispagestyle{empty}

\vspace*{-.5cm}
\begin{center}
Department of Computer Science\\University of Texas - Rio Grande Valley \\Edinburg, TX 78539-2999, USA\\
\{robert.schweller, timothy.wylie\}@utrgv.edu
\end{center}

\begin{abstract}
In this paper, we seek to provide a simpler proof that the relocation problem in Ricochet Robots (Lunar Lockout with fixed geometry) is PSPACE-complete via a reduction from Finite Function Generation (FFG). Although this result was originally proven in 2003, we give a simpler reduction by utilizing the FFG problem, and put the result in context with recent publications showing that relocation is also PSPACE-complete in related models. 
\end{abstract}


\section{Introduction}

\begin{figure}[t]
    \centering
	\begin{subfigure}[b]{0.4\textwidth}
        \centering
        \includegraphics[height=2.in]{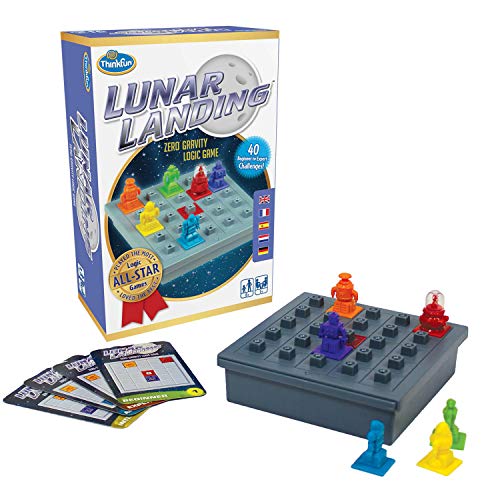}
        \caption{Lunar Lockout}\label{fig:ll}
    \end{subfigure}
	\begin{subfigure}[b]{0.45\textwidth}
        \centering
        \includegraphics[height=1.9in]{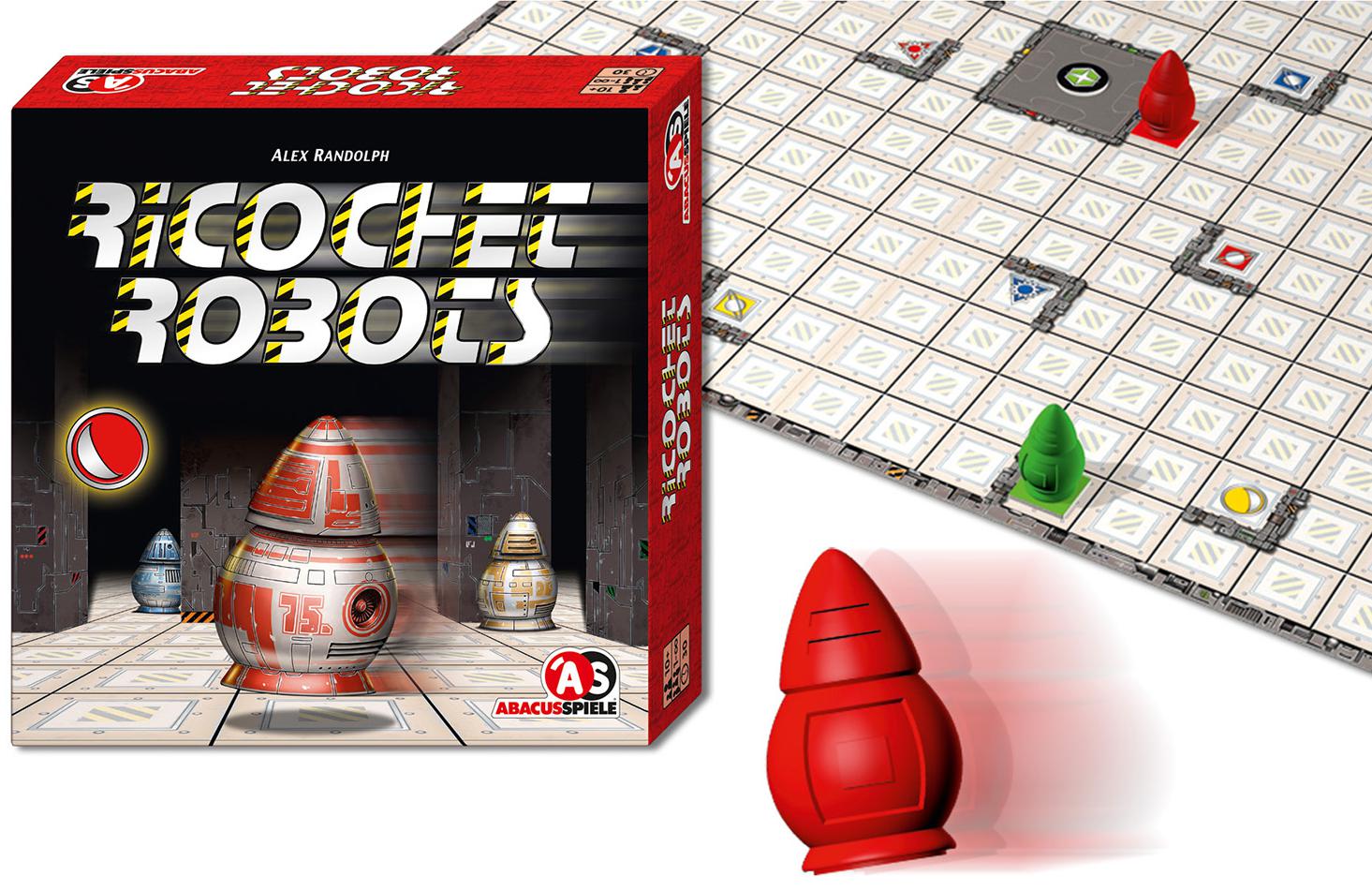}
        \caption{Ricochet Robots}\label{fig:rr}
    \end{subfigure}
    \caption{The board games Lunar Lockout and Ricochet Robots.}
    \label{fig:games}
\end{figure}

In this paper we study the complexity of relocation within the puzzle-game Ricochet Robots~\cite{Ricochet}, which is equivalent to the game Lunar Lockout with fixed geometry \cite{LunarLockout} (see Figure \ref{fig:games} for pictures). The Ricochet Robots puzzle consists of a 2D grid \emph{board} containing polyomino obstacles and a collection of unit-size robots placed on the board.  The player may select any robot and move it maximally in any of the four cardinal directions.  With this basic operation, the goal (relocation problem) is to move a target robot to a target goal location on the board.

While deceptively simple, even the single-player puzzle version of this game has proven to be quite complex, with online solvers being written to help develop solving strategies~\cite{Scherer:solver,Cox:solver,Noel:solver}. 
The relocation problem with Lunar Lockout was originally shown to be NP-hard in 2001 in \cite{Hock:2001:MS}. With fixed geometry, it was shown to be PSPACE-complete in 2003 in \cite{Hartline:2003:SIAM}. Recent work has also focused on some parameterized results for Ricochet Robots \cite{Hesterberg:2017:JIP}.

The proof of PSPACE-hardness was proven by showing any polynomial-space Turing machine can be transformed into an instance of Lunar Lockout with fixed geometry \cite{Hartline:2003:SIAM}. In this paper, we show a simpler proof by reducing from Finite Function Generation \cite{kozen1977lower}, which was proven to be PSPACE-complete in 1977. 

This work fits into a larger landscape of ``Tilt" problems that have received recent interest.  Given a 2D board with both open locations and blocked locations, as well as a set of unit-size robots placed at open locations on the board, the question of relocating a particular robot to a particular location has been studied under two fundamental variants:  \emph{global} signals versus \emph{local} signals, and \emph{unit steps} versus maximal \emph{full steps}.  In the case of global signals, each move (in one of the four cardinal directions) moves ALL robots on the board in the specified direction.  In terms of step distance, in the \emph{unit step} variant, a moved robot takes just a single step in the specified direction, while in the \emph{full step} variant each robot moves maximally in the specified direction until a wall or another robot is encountered.  Together, these variants create four natural versions of the relocation problem for consideration.

In the case of global signals and full steps, the relocation problem was recently shown to be PSPACE-complete in SODA 2020~\cite{FullTiltSequel}.  In the case of global signals with single-steps, the problem has also recently been shown to be PSPACE-complete~\cite{Caballero:2020:CCCG1}.  For local signals and single-steps, the problem is easily solved in polynomial time.  However, recent work by Brunner, Chung, Demaine, Hendrickson, Hesterberg, Suhl, and Zeff~\cite{Brunner:2020:ARXIV} have considered an interesting variant in which each piece has a provided path that it must travel along, making the problem PSPACE-complete.  Within this landscape, the  remaining of the four natural tilt models is Ricochet Robets with local signals and full steps.  A summary of these results is provided in Table~\ref{tab:overview}.


%
%
%


\begin{table*}[t]
    \centering 
    \def\arraystretch{1.3}
    \begin{tabular}{|  c | c | c | c | c |} 
    \hline
        \textbf{Tilt Model} & \textbf{Signal} & \textbf{Polyominoes} & \textbf{Complexity} & \textbf{Reference} \\ \hline
         {\multirow{3}{*}{Unit}} & Single & $1 \times 1$ & P &  \\ \cline{2-5}
          & Single (paths) & $1 \times 1$ & PSPACE-complete &  \cite{Brunner:2020:ARXIV}\\ \cline{2-5}
          & Global & $1 \times 1$ & PSPACE-complete &  \cite{Caballero:2020:CCCG1}\\ \hline
         {\multirow{2}{*}{Full}} & Single & $1 \times 1$ & PSPACE-complete &   \cite{Hartline:2003:SIAM}, Thm. \ref{thm:ricochet}\\ \cline{2-5}
         & Global & $1 \times 1$ & PSPACE-complete &  \cite{FullTiltSequel}\\ \hline
    \end{tabular}
    \caption{An overview of related models and the complexity of the relocation problem. All have concrete or fixed polyominos that can not move in a connected board. The Single (paths) model requires that every tile moves along a specific path given for each tile. This paper gives Theorem \ref{thm:ricochet}, which proves relocation in full tilt with single signaling.}\label{tab:overview}
\end{table*}
\section{Model Preliminaries} \label{Prelims}

\para{Board}
A \emph{board} (or \emph{workspace}) is a rectangular region of the 2D square lattice in which specific locations are marked as \emph{blocked}.  Formally, an $m\times n$ board is a partition $B=(O,W)$ of $\{(x,y) | x\in \{1, 2, \dots, m\}, y\in \{1, 2, \dots, n\}\}$ where $O$ denotes a set of \emph{open} locations, and $W$ denotes a set of \emph{blocked} locations- referred to as ``concrete'' or ``walls.''  Based on a geometric hierarchy \cite{FullTiltSequel}, here we create a \emph{connected} board\footnote{Note that within this model, every board geometry could be considered as rectangular and the blocked spaces are simply robots that are never given a move signal. However, for the reduction, these must be non-movable robots and thus we adhere to the definition of a blocked space instead.}. A board is said to have \emph{connected} geometry if the set of open spaces $O$ for a board is a connected shape.

\para{Tiles}
A tile is a labeled unit square centered on a non-blocked point on a given board. Formally, a tile is an ordered pair $(c,a)$ where \emph{c} is a coordinate on the board, and \emph{a} is a tile label. In this work we have no attachments between tiles and simply use it as an identifying label. 

\para{Configurations}
A configuration is an arrangement of tiles on a board such that there are no overlaps among tiles, or with blocked board spaces.  Formally, a configuration $C=(B, P=\{P_1,\ldots, P_k\})$ consists of a board $B$, along with a set of non-overlapping tiles $P$ that each do not overlap with the blocked locations of board $B$.

\para{Particle Step} A \emph{particle step} is a way to turn one configuration into another by way of a signal that moves a tile $t$ in a configuration one unit in a direction $d \in \{N,E,S,W\}$ when possible without causing an overlap with a blocked location or another tile. If a configuration does not change under a step transition for tile $t$ in direction $d$, we say the configuration is \emph{$d(t)$-terminal}. 


\begin{figure}[t]
    \centering
	\begin{subfigure}[b]{0.115\textwidth}
        \includegraphics[width=1.\textwidth]{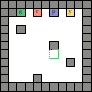}
        \caption{Start}
    \end{subfigure}
	\begin{subfigure}[b]{0.115\textwidth}
        \includegraphics[width=1.\textwidth]{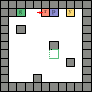}
        \caption{E(r) }
    \end{subfigure}
    \begin{subfigure}[b]{0.115\textwidth}
        \includegraphics[width=1.\textwidth]{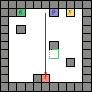}
        \caption{S(r)}
    \end{subfigure}
    \begin{subfigure}[b]{0.115\textwidth}
        \includegraphics[width=1.\textwidth]{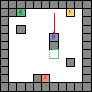}
        \caption{S(p)}
    \end{subfigure}
        \begin{subfigure}[b]{0.115\textwidth}
        \includegraphics[width=1.\textwidth]{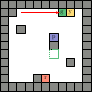}
        \caption{E(g)}
    \end{subfigure}
        \begin{subfigure}[b]{0.115\textwidth}
        \includegraphics[width=1.\textwidth]{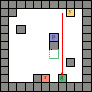}
        \caption{S(g)}
    \end{subfigure}
        \begin{subfigure}[b]{0.115\textwidth}
        \includegraphics[width=1.\textwidth]{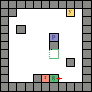}
        \caption{W(g)}
    \end{subfigure}
    \begin{subfigure}[b]{0.115\textwidth}
        \includegraphics[width=1.\textwidth]{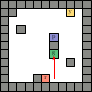}
        \caption{N(g)}
    \end{subfigure}
    \caption{Ricochet Robots relocation example. Given the starting configuration (Start), a specific tile (green tile g), and a specific location (outlined), this provides a sequence of particle tilts that places g in the location. The tiles are labeled g, r, p, y based on their respective colors of green, red, purple, and yellow. The grey tiles are all blocked locations that can not be moved.}
    \label{fig:simple_example}
\end{figure}


\para{Particle Tilt} A \emph{particle tilt} in direction $d \in \{N,E,S,W\}$ for a configuration is executed by repeatedly applying a particle step in direction $d \in \{N,E,S,W\}$ on the same tile $t$ until a $d(t)$-terminal configuration is reached. We denote this as $d(t)$. We say that a configuration $C$ can be \emph{reconfigured in one move} into configuration $C'$ (denoted $C \rightarrow_1 C'$) if applying one particle tilt on a tile $t$ in some direction $d$ to $C$ results in $C'$.  We define the relation $\rightarrow_*$ to be the transitive closure of $\rightarrow_1$. Therefore, $C \rightarrow_* C'$ means that $C$ can be reconfigured into $C'$ through a sequence of particle tilts.


\para{Particle Tilt Sequence} A \emph{particle tilt sequence} is a sequence of particle tilts. For a given tile, the sequence can be inferred from a series of directions $ D = \langle d_1, d_2,\dots, d_k \rangle$; each $d_i \in D$ implies a particle tilt in that direction on a tile. For simplicity, when discussing a particle tilt sequence on a specific tile, we just refer to the series of directions from which that sequence was derived. Given a starting configuration, a particle tilt sequence corresponds to a sequence of configurations based on the tilt transformation.  An example particle tilt sequence $\langle E(r), S(r), S(p), E(g), S(g), W(g), N(g) \rangle$ and the corresponding sequence of configurations can be seen in Figure \ref{fig:simple_example}.


\para{Ricochet Robots} For this paper, we refer to this tilt model as Richochet Robots for convenience.

\para{Finite Function Generation} \label{p:ffg}
In our result we make use of \emph{finite function generation} which was shown to be PSPACE-complete in \cite{kozen1977lower}.
Let $A$ be a set with $N$ elements, $F = \{ f_1, \dots, f_k \}$ be a finite set of maps $f_i : A \to A$, and $h: A \to A$. Is $h$ generated by
some sequence of compositions $s = f_i \circ \dots \circ f_n\ s.t.\ \{ f_i, \dots, f_n \} \in F$? More succinctly, $FFGEN = \{ F, s, h \}\  |  h $  is  generated  by some sequence of compositions $s$ whose elements exist in F$\}$. 

\begin{figure}[t]
    \centering
	\begin{subfigure}[b]{.27\textwidth}
        	\includegraphics[width=.9\textwidth]{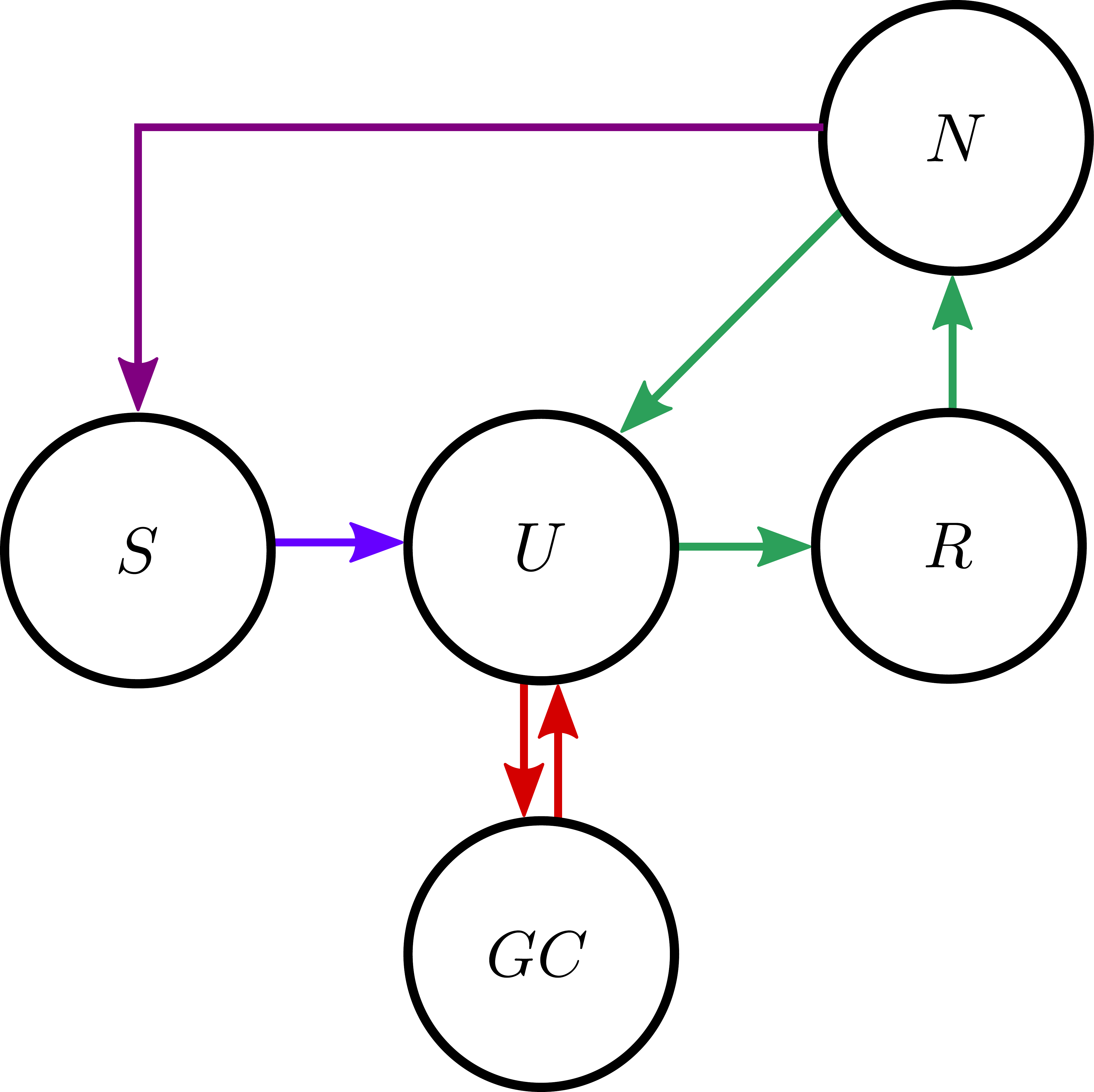}
		\caption{}
	\end{subfigure} \hspace*{.5cm}
	\begin{subfigure}[b]{.5\textwidth}
		\centering 
                \def\arraystretch{1.1}
        		\begin{tabular}{|p{1.3cm}|p{7cm}|}\hline
    			\textbf{Com.} & \textbf{Description} \\ \hline
    			1. $S$ & Select function for the first element in 
			        the domain.\\ \hline
			2. $U$ & Unlock next element. \\ \hline
    			3. $R$ & Input the element to its lock gadget.\\ \hline
    			4. $N$ & Input unlocked element to the function 
			        determined by previous element. \\ \hline
			5. $G$ & Input element to the goal gadget 
			        and move to next element. \\ \hline
			\end{tabular}
		\caption{}
	\end{subfigure}
	\caption{(a) Flowchart of the relocation process. Purple paths lead to a state that can only be reached with the last domain element. Blue paths lead to states that can only be reached with the first domain element. Red paths can be traversed only once per element. (b) Descriptions of each state in the flowchart.} \label{fig:flowchart}
\end{figure}

\begin{figure}[t]
    \centering
        \begin{subfigure}[b]{.47\textwidth}
                \includegraphics[width=1\textwidth]{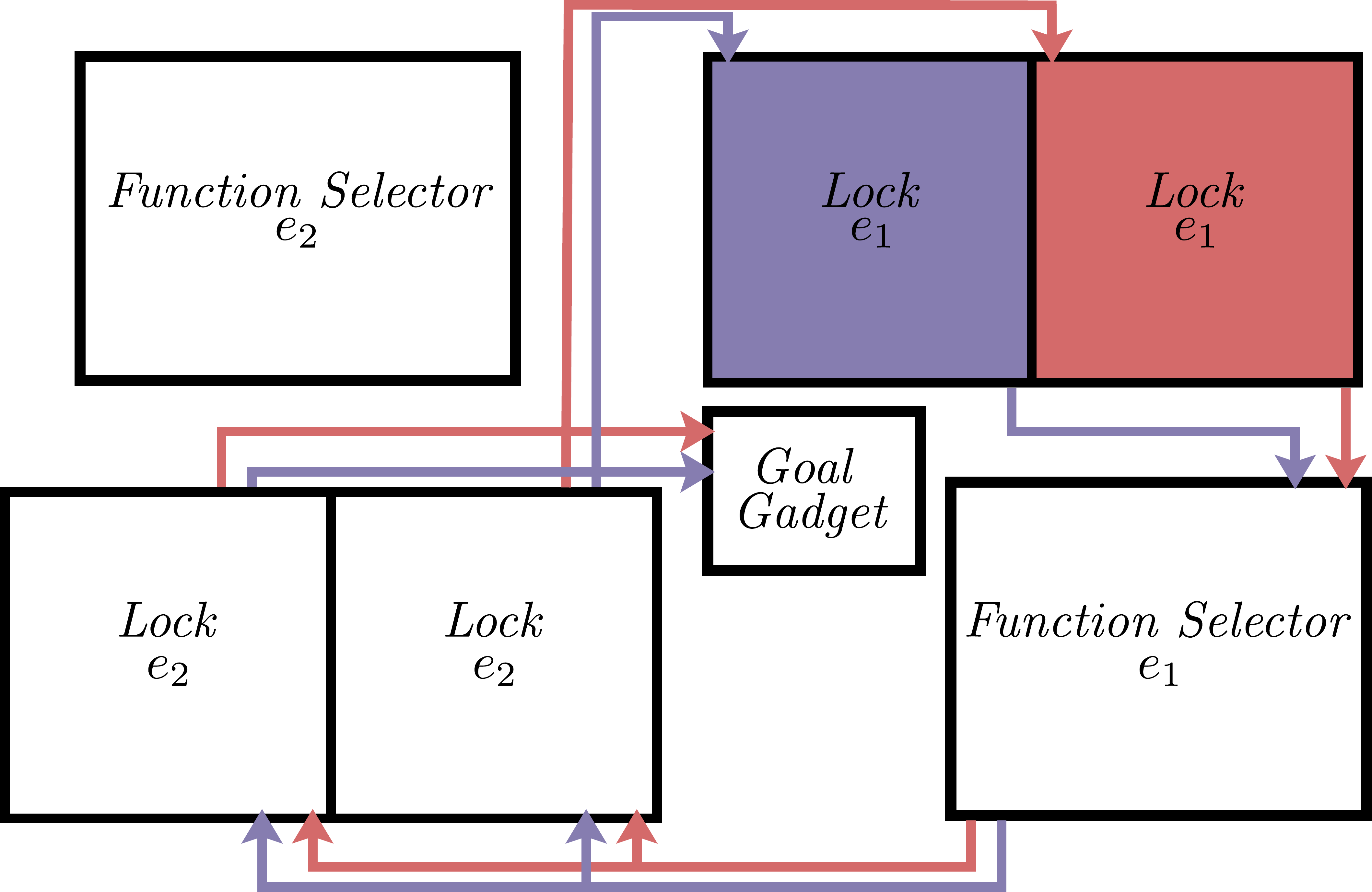}
                \caption{}
                \label{fig:ffgrr_example_e1}
        \end{subfigure} 
        \begin{subfigure}[b]{.47\textwidth}
                \includegraphics[width=1\textwidth]{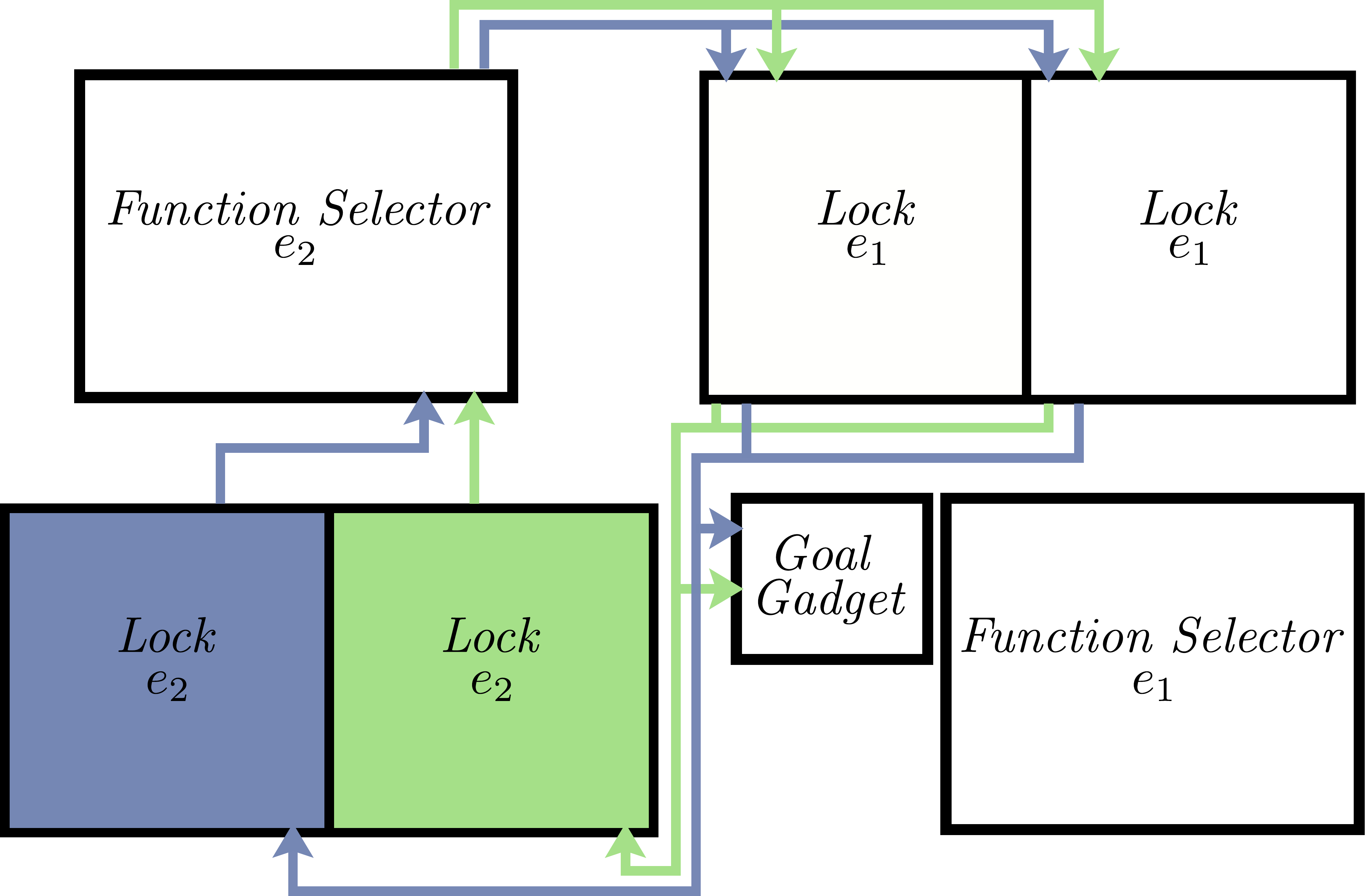}
                \caption{}
                \label{fig:ffgrr_example_e2}
        \end{subfigure}
        \caption{Example of a two element system. 
	}
        \label{fig:fw_example}
\end{figure}

\begin{figure}[t]
    \centering
	\begin{subfigure}[b]{.47\textwidth}
		\includegraphics[width=1\textwidth]{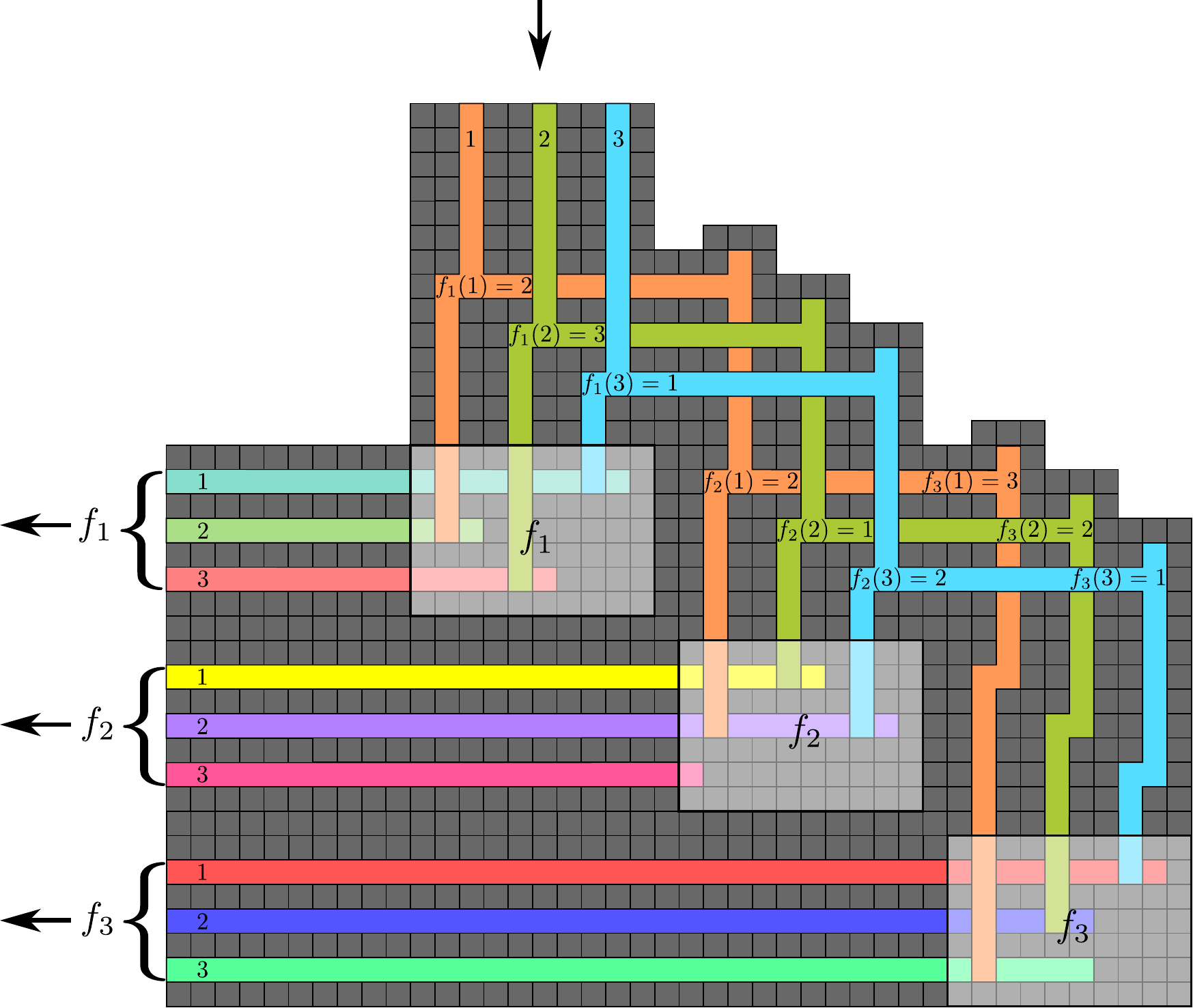}
		\caption{Function Selector}
		\label{fig:fs_example_e1}
    \end{subfigure}	
    \hspace*{.5cm}
	\begin{subfigure}[b]{.47\textwidth}
		\includegraphics[width=1\textwidth]{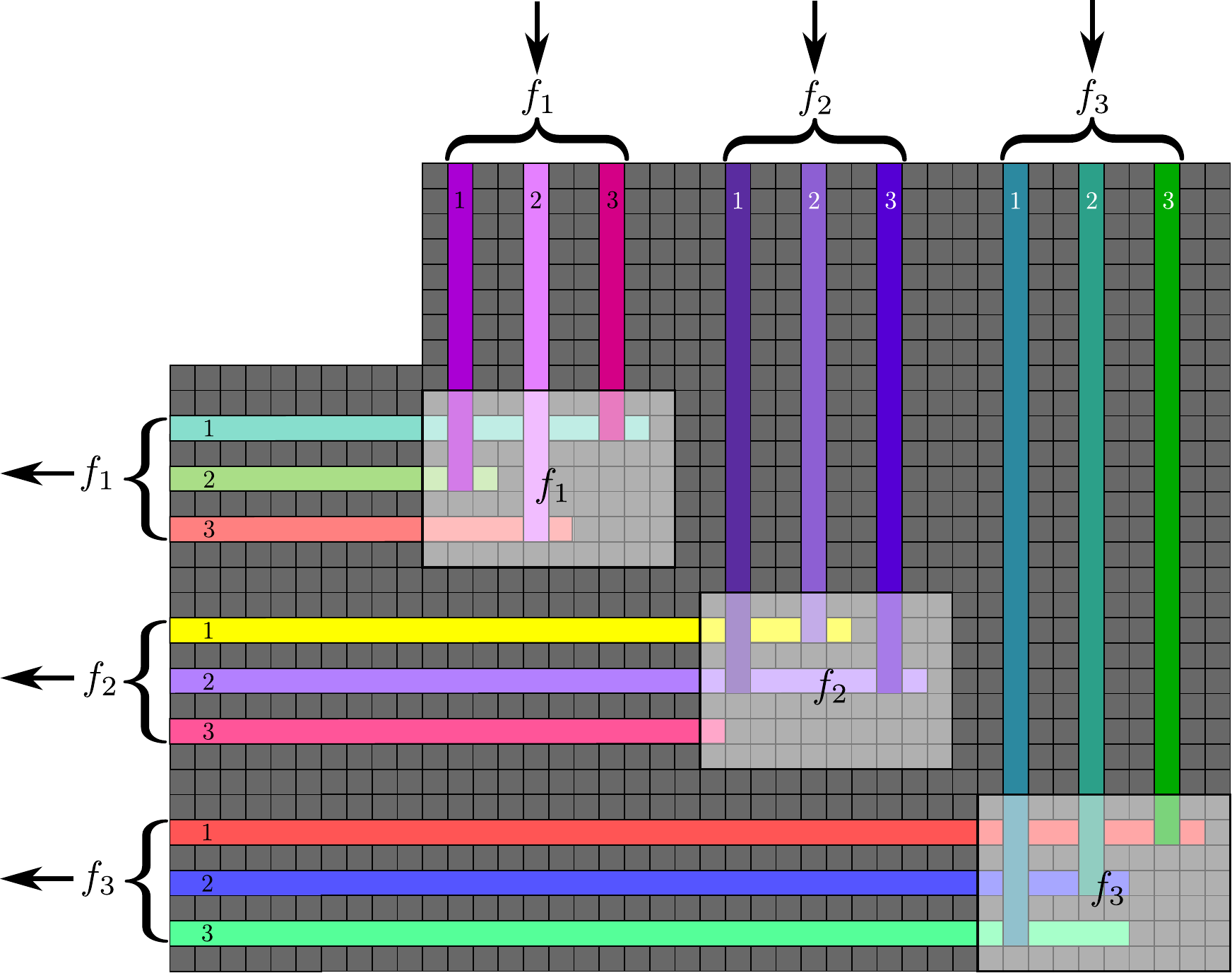}
		\caption{Function Enforcer}
		\label{fig:fs_example_ei}
	\end{subfigure}

	\caption{An example of function selector and function enforcer gadgets with domain $D = \{1,2,3\}$ and functions $f_1,f_2, f_3\ s.t.\ f_1(1) = 2, f_1(2) = 3, f_1(3) = 1, f_2(1) = 2, f_2(2) = 1, f_2(3) = 2, f_3(1) = 3, f_3(2) = 2, f_3(3) = 1$. The arrows represent the possible input and output of the gadgets.}
	\label{fig:fs_example}
\end{figure}

\begin{figure}[t]
	\centering
 \vspace*{-1.5cm}
	\begin{subfigure}[b]{.1\textwidth}
        \centering
        \includegraphics[width=1\textwidth]{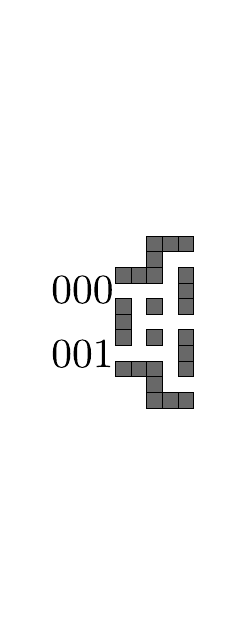}
        
        \vspace{-2.5cm}
        
		\includegraphics[width=1\textwidth]{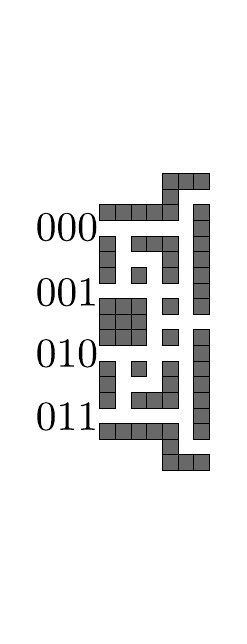}
        \vspace{-1.3cm}
		\caption{}
		\label{fig:ls_example_4}
	\end{subfigure} 
	\begin{subfigure}[b]{.1\textwidth}
		\includegraphics[width=1\textwidth]{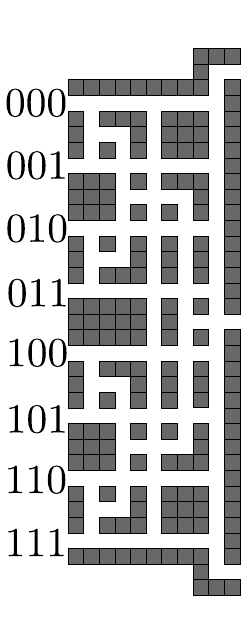}
		\caption{}
		\label{fig:ls_example_8}
	\end{subfigure}
    	\begin{subfigure}[b]{.39\textwidth}
		\centering
        	\includegraphics[width=.9\textwidth]{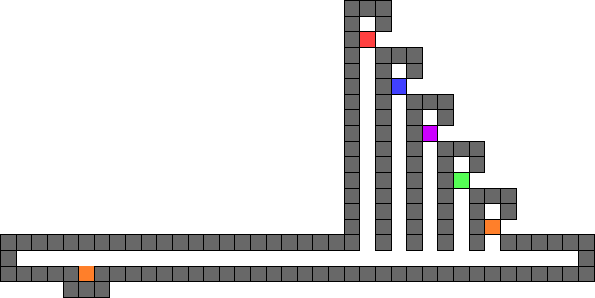}
		\caption{Relocation Goal}\label{fig:relocgoal}
	\end{subfigure}
	\begin{subfigure}[b]{.39\textwidth}
		\centering
        	\includegraphics[width=.9\textwidth]{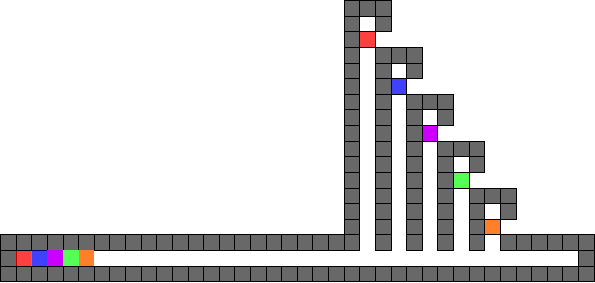}
        	\caption{Reconfiguration Goal}\label{fig:recongoal}
	\end{subfigure}
	\caption{(a-b) An example of lock selector gadgets of a system with domain of size 2, 4, and 8. (c) Goal gadget for relocation. (d) Goal gadget for reconfiguration.}
	\label{fig:ls_example}\label{fig:goal_gadgets}
\end{figure}

\section{Complexity of Ricochet Robots}

\begin{figure}[t]
    \centering
    \begin{subfigure}[b]{.37\textwidth}
		\includegraphics[width=1\textwidth]{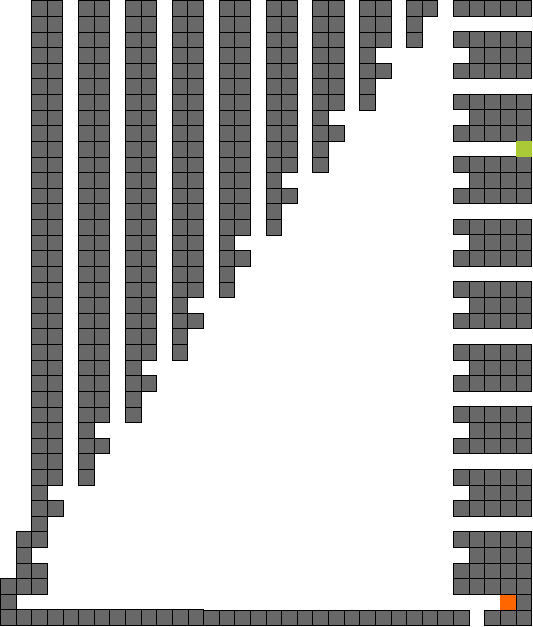}
		\caption{}
		\label{fig:l_path}
	\end{subfigure}
    \hspace{.7in}
    \begin{subfigure}[b]{.4\textwidth}
		\includegraphics[width=1\textwidth]{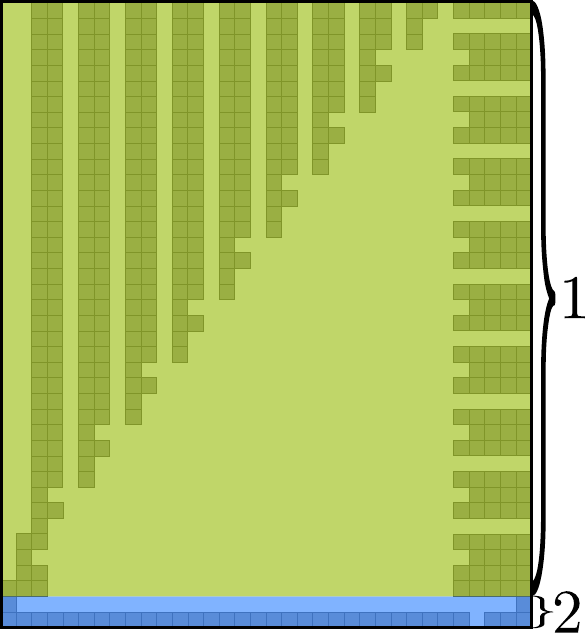}
		\caption{}
		\label{fig:l_paths}
	\end{subfigure}   
	\caption{(a) Lock Gadget with labeled functions and sample imputs. The paths of the locked tile(orange) and the unlocking tile (light green). The unlocked tile cannot complete its traversal without the unlocking tile. (b) Lock Gadget with labeled sections. Section one is the area where the unlocking tile can exist. Section two is the intersection between section one and two, and is the area where the unlocking tile and the locked tile interact. It is also the area where the locked tile can exist.}
	\label{fig:misc}
\end{figure}

Here, we give a reduction from \emph{finite function generation} in which, given $m$ functions $f_i: X \to X$ where $|X|$ is polynomial in size and $1 \leq i \leq m$, and a function $g(x): X \to X$, is $g$ the composition of some sequence of $f_i$'s? We first discuss the reduction framework in detail.

\subsection{Reconfiguration Framework for Individually Controlled Particle Systems}\label{s:ffgfw}

We will employ a framework Figure \ref{fig:flowchart} that will force the elements in the system through the same function, else $h$ cannot be generated. This framework works by selecting a function and inputting the first element through it with a \emph{function selector} gadget. After this, we will need the element to ``unlock" the next element in the domain. This next element cannot be input into its function if it is not unlocked first. Attempting to force an element to its function selector without unlocking it first would render the system not reconfigurable. We achieve this with the use of \emph{lock} gadgets which are also used to determine which function the unlocked element will be input through. This process is then repeated with all the elements in the system. Once all the elements have been input through a function, and the first element has been unlocked by the last, it is used to determine which function will follow next. This forces that all element to be input through the same function the same number of times. The elements can only be input into a \emph{goal gadget} after unlocking the next element, but before being input into their lock gadget. This ensures that all the tiles will be input through the same function before being input to the goal gadget.

\begin{figure}[t]
    \centering
    \begin{subfigure}[b]{.37\textwidth}
		\includegraphics[width=1\textwidth]{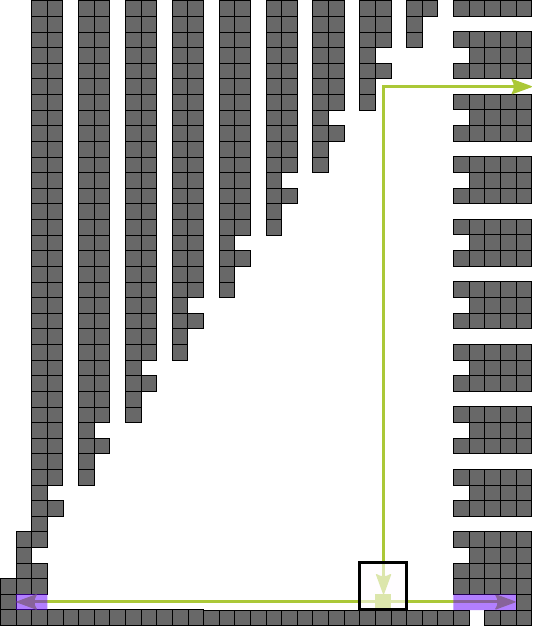}
		\caption{}
		\label{fig:lemma_unlock}
	\end{subfigure}
    \hspace{.7in}
    \begin{subfigure}[b]{.37\textwidth}
		\includegraphics[width=1\textwidth]{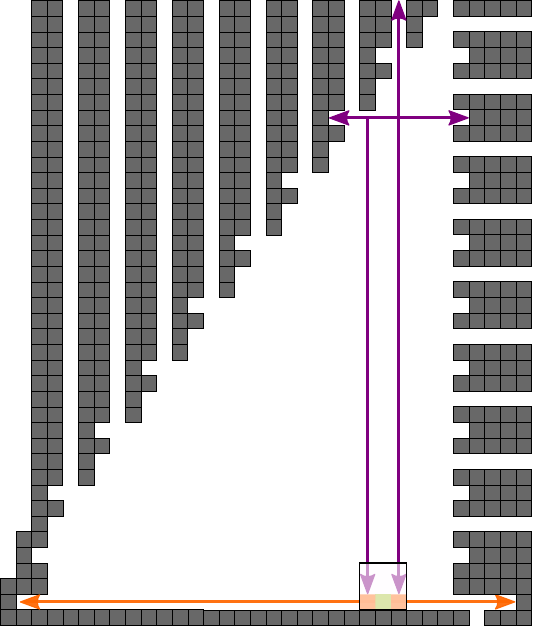}
		\caption{}
		\label{fig:lemma_locked}
	\end{subfigure}
    \caption{ (a) Possible paths for an unlocking tile. The green paths indicate the paths the locked tile can reach by itself. The blue areas indicate when a tile becomes stuck. The highlighted section is where tiles are meant to interact. (b) Possible paths for an unlocking tile. The highlighted section is where tiles can interact. The orange paths indicate the paths the locked tile can reach by itself. Purple paths indicate the paths the locked tile can reach with the help of the unlocking tile. The blue areas indicate when a tile becomes stuck.}
    \label{fig:misc2}
\end{figure}

\begin{figure*}[t]
    \vspace{-.2cm}
    \centering 
     \textbf{Function Selector and Function Enforcers Sequences}
     \def\arraystretch{1.1}
	\begin{tabular}{|p{.9cm}|p{4.cm}|p{9cm}|}\hline 
		\textbf{Name} & \textbf{Tilt Sequence} & \textbf{Description} \\ \hline
		$SF$ & $\langle S \rangle + {\langle E, S \rangle}^i + \langle W, S, W \rangle$ \rule{0pt}{2.6ex} & Selects the function all the elements the domain will be input to. $i$ specifies the function to be selected. \\ \hline
		$EF$ & $\langle S, E \rangle$ \rule{0pt}{2.6ex} & Enforces all elements are input through the same function as the first element. \\ \hline
    \end{tabular}
    
    \vspace{.2cm}
    \textbf{Lock Selector Sequences}
    
    \begin{tabular}{|p{.9cm}|p{4.cm}|p{9cm}|}\hline 
		$T_0$ & $\langle W, N \rangle$ \rule{0pt}{2.6ex} & Selects $0$ for that bit position. \\ \hline 
		$T_1$ & $\langle W, S \rangle$ \rule{0pt}{2.6ex} & Selects $1$ for that bit position. \\ \hline 
		$RF$ & $\langle N, E \rangle$ \rule{0pt}{2.6ex} & Inputs tile to function selector or function enforcer gadget after returning from lock gadget. \\ \hline 
		$G$ & $\langle S, E \rangle$ \rule{0pt}{2.6ex} & Inputs tile to goal gadget after returning from lock gadget. \\ \hline 
    \end{tabular}

    \vspace{.2cm}
    \textbf{Lock Sequences}
    \begin{tabular}{|p{.9cm}|p{4.cm}|p{9cm}|}\hline 
		$P$ & $\langle W, S \rangle$ \rule{0pt}{2.6ex} & Positions unlocking tile for unlock. \\ \hline 
		$U$ & $\langle W, N \rangle$ \rule{0pt}{2.6ex} & Unlocks locked tile. \\ \hline
		$R$ & $\langle N, E \rangle$ \rule{0pt}{2.6ex} & Returns tile to lock selector. \\ \hline
	\end{tabular}
	\caption{Particle Tilt Sequences.}
	\label{tab:sequences}
\end{figure*}

As an example, we can refer to Figure \ref{fig:fw_example}. Here, $e_1$ is located in either of its two lock gadgets and the element is not locked. We perform step \emph{S} by inputting $e_1$ to its function selector gadget and choosing a function. Once the function has been chosen and $e_1$ has been mapped to its new value, we perform $U$ and unlock $e_2$. For that we need to choose which lock gadget $e_2$ is located at with the use of a \emph{lock selector} gadget which can be seen in the appendix. Once we have unlocked $e_2$ we perform $R$ and return $e_1$ to its lock gadget. Now we can perform $N$ and input $e_2$ to its function selector, and perform $U$ and $R$ respectively to unlock $e_1$ and return $e_2$ to its lock gadget. Since $e_2$ is the last gadget, we can now perform $S$ with $e_1$ again. However, if we wanted to input the elements to the goal gadget, we would do so before returning the element to its lock gadget. 

For each element in the domain $A$, there is one function selector gadgets and $|A|$ lock gadgets. 
Each lock gadget represents an element's possible value. Since an element might match to any of the domain inputs, there must be a lock gadget for each domain value. 
Therefore, an element's lock gadget represents the element's value and is mapped to the correct input in its function selector. However, once an element has been output through a function selector, it has to be relocated to the lock gadget where the next element resides. Since the next element could reside in any of its lock gadgets we utilize a \emph{lock selector} gadget to choose between the next element's lock gadgets. There are $|F| \times |A|$ lock selectors, one for every element in every function. This gadget connects the element's function selector or enforcer, lock gadgets and the next element's lock gadget.



\subsection{Reconfiguration Under Ricochet Robots is PSPACE-Complete}\label{s:ffgrr}
Ricochet Robots systems use particle tilts, and motion is not universal. 
Formally, given two configurations $C = (B, P )$ and $C_{success} = (B, P')$, does there exist a tilt sequence such that $C \to_* C_{success}$? More formally, to represent an $FFGEN$ system we will use a set of configurations $A_C$ and let $F_C\ =\ \{ {f_C}_1, \dots, {f_C}_k \}$ be a set of maps ${f_C}_i : A_C \to A_C$,  $h_C: A_C \to A_C$, and $ s_C = {f_C}_i \circ \dots \circ {f_C}_n\ s.t.\ \{{f_C}_i, \dots, {f_C}_n\} \in F_C $. An initial configuration $C$ can be reconfigured into $C_{success}$, iff $h_C(C) = C_{success}$ and $\ s_C(C) = C_{success}$. If the above is not true, the system would result in a terminal configuration $C_{fail}$ where $C{fail}\not\to_* C_{success} $. The tilts used to move the tile through the function selector and function enforcer gadgets can be seen as an element being input through different functions. One would be able to reconfigure the system to the goal configuration if and only if $h_C$ was generated by some sequence ${f_C}_i \circ \dots \circ {f_C}_n$.
Finally, we can be more specific and talk about tiles being input and output into gadgets, since moving tiles and changing configurations is equivalent. However, it is easier to explain some gadgets when talking about tiles as input, instead of configurations.

\subsubsection{Function Selector}
Our function selection gadget takes a tile as input through a \emph{mapping tunnel} that dictates the input value to that function, and through a series of particle tilts selects the function all the tiles will be input through. It will output the tile through a mapping tunnel that indicates the function and the resulting value. 
This gadget has $|A|$ input mapping tunnels to represent all the possible values an element can have and $|F| \times |A|$ output mapping tunnels to represent the possible values and to indicate the function for the rest of the elements. 
Once a tile has been output, it cannot enter the function selector again through the output mapping tunnel.
An example is seen in Figure \ref{fig:fs_example}, we can see how elements are mapped to others and a function is selected when being input through the function selector.  
The Particle Tilt Sequence for traversal of this gadget is $SF$ found in Figure \ref{tab:sequences}.


\subsubsection{Function Enforcer}
A \emph{function enforcer} gadget is responsible for enforcing the rest of the domain elements are input through the same function. For this gadget there are $|F| \times |A|$ mapping tunnels, since it is necessary to preserve the function chosen by the function selector. It similarly as the function selector, except a function cannot be chosen.
The Particle Tilt Sequences for traversal of this gadget is $EF$ found in Figure \ref{tab:sequences}. 


\subsubsection{Lock Selector}
Once an element exits through a function selector it will be input to one of $|A|$ \emph{lock selectors}. The purpose of these selectors is to input the element to the next element's lock gadget. Since there are $|A|$ lock gadgets, which preserve the value of its domain element, the element can be located in any of these lock gadgets. Once you have unlocked the next element, the current element has to be input back to the lock selector to reach its own lock gadget. The lock selectors also preserve the value of the element by inputting it to the lock gadget that corresponds to the correct value. An example of lock selectors for different domain sizes can be seen in Figure \ref{fig:ls_example}.

\subsubsection{Lock Gadget}
The \emph{lock} gadget is essential in the construction of the proof because it forces the input of all elements through the same function. Its input is two tiles, one for the element to be unlocked and one for the proceeding element performing the unlock. The first tile enters the gadget from an \emph{enforcer tunnel} and must be located in the gadget before the second tile enters the gadget through one of $|F| \times |A|$ enforcer tunnels. The enforcer tunnels ensure that all tiles input and output from the gadget maintain their value and dictates what function the unlocked tile will be input to. The locked tile can only move through the gadget if and only if the unlocking tile is in the corresponding position. If not intended particle tilts are performed the tiles will become trapped and will never be able to leave the lock gadget. 

\begin{lemma} \label{lemma:unlock}
        An unlocking tile can only exit a lock gadget through its input mapping tunnel.
\end{lemma}
We prove this by showing all the possible paths an unlocking tile can take inside a lock gadget, and showing that any particle tilt sequence for the unlocking tile that is not defined in Figure \ref{tab:sequences} either makes the unlocking tile stuck or does not affect the gadget's output. Figure \ref{fig:lemma_unlock} shows all the possible paths an unlocking tile could take. Since the unlocking cannot use the locked tile to reach any positions it is not supposed to reach, any unexpected particle tilt sequence would lead to a position where the unlocking tile gets stuck. These positions are indicated in section two of Figure \ref{fig:lemma_unlock}.

\begin{lemma}
        A locked tile can only exit a gadget through its output mapping tunnel.
\end{lemma}

We prove this similarly to \ref{lemma:unlock} by showing all the possible paths a locked tile can take inside a lock gadget, and showing that any particle tilt sequence for the locked tile that is not defined in Figure \ref{tab:sequences} either makes the locked tile stuck or does not affect the gadget's output. Figure \ref{fig:lemma_locked} shows all the possible paths for a locked tile. The locked tile can only interact with the unlocking tile in one section, and there is only one path that would let the locked tile leave the gadget.


\subsubsection{Relocation Goal Gadget}

The relocation goal gadget is rather simple (see Figure \ref{fig:relocgoal}). It consists of the goal for the last robot in the domain, and ensures that the robot must be moved into its goal area last. The theorem follows from Lemmas \ref{thm:ricochethard} and \ref{thm:ricochetmemb}, and Lemma \ref{thm:ricochethard} follows from the correctness of the presented construction. 

\begin{lemma}\label{thm:ricochethard}
Relocation within Ricochet Robots with fixed polyominoes and $1\times 1$ movable polyominoes is PSPACE-hard \cite{Hartline:2003:SIAM}.
\end{lemma}

\begin{lemma}\label{thm:ricochetmemb}
Relocation within Ricochet Robots with fixed polyominoes and $1\times 1$ movable polyominoes is in PSPACE \cite{Hartline:2003:SIAM}.
\end{lemma}

\begin{theorem}\label{thm:ricochet}
Relocation within Ricochet Robots with fixed polyominoes and $1\times 1$ movable polyominoes is PSPACE-complete \cite{Hartline:2003:SIAM}.
\end{theorem}


\subsubsection{Reconfiguration Goal Gadget}

The reconfiguration goal gadget is almost the same as the reconfiguration gadget (see Figure \ref{fig:recongoal}). It consists of the goal areas for all the robots in the domain, and ensures that the robots must be moved into their goal areas in the order that they appear in the domain.

\begin{corollary}\label{cor:ricochet}
Reconfiguration within Ricochet Robots with fixed polyominoes and $1\times 1$ movable polyominoes is PSPACE-complete.
\end{corollary}

\section{Conclusion}
In this paper we gave a simpler reduction to show that relocation and reconfiguration in the tilt model with addressable robots (Ricochet Robots or Lunar Lockout with fixed blocks) is PSPACE-complete even when all particles are $1 \times 1$ tiles. This question is still open if there is no fixed geometry (or fixed robots). Is there a way to simulate some fixed geometry with other pieces?

\bibliographystyle{amsplain}
\bibliography{tilt}


\end{document}